\documentclass[showpacs]{revtex4}
\usepackage{graphicx}
\usepackage{dcolumn}
\usepackage{amsmath}
\usepackage{color}
\usepackage{epstopdf}
\usepackage{graphicx}
\usepackage{subfigure}
\usepackage[utf8x]{inputenc}
\definecolor{coolblack}{rgb}{0.0, 0.18, 0.39}
\definecolor{darkred}{rgb}{0.5,0,0}
\definecolor{darkgreen}{rgb}{0,0.5,0}
\definecolor{darkblue}{rgb}{0,0,0.5}
\definecolor{lapislazuli}{rgb}{0.15, 0.38, 0.61}
\definecolor{venetianred}{rgb}{0.78, 0.03, 0.08}
\definecolor{bleudefrance}{rgb}{0.19, 0.55, 0.91}
\definecolor{dogwoodrose}{rgb}{0.84, 0.09, 0.41}
\usepackage[linktocpage,colorlinks]{hyperref}
\hypersetup{colorlinks=true, citecolor=darkgreen, linkcolor=blue,urlcolor = darkblue}
\makeatletter
\def\btt#1{\texttt{\@backslashchar#1}}
\DeclareRobustCommand\bblash{\btt{\@backslashchar}} \makeatother

\begin{document}
\title{Deflection of Light and Shadow Cast by a Dual Charged Stringy Black Hole}
\author{Shubham Kala $^{a}$}\email{shubhamkala871@gmail.com}
\author{Saurabh $^{b}$}\email{sbhkmr1999@gmail.com}
\author{Hemwati Nandan$^{a,c}$}\email{hnandan@associates.iucaa.in}
\author{Prateek Sharma $^{a}$}\email{prteeksh@gmail.com}
\affiliation{$^{a}$Department of Physics, Gurukula Kangri Vishwavidyalaya, Haridwar 249 404, Uttarakhand, India}
\affiliation{$^{b}$Department of Physics, Dyal Singh College, University of Delhi, New Delhi 110003, India}
\affiliation{$^{c}$Center for Space Research, North-West University, Mahikeng 2745, South Africa}

\begin{abstract}
	\noindent
Gravitational lensing and black hole shadows are one of the strongest observational evidences to prove the existence of black holes in the universe. The gravitational lensing arises due to the deflection of light by the gravitational field of a gravitating body such as a black hole. Investigation of the shadow cast by a compact object as well as deflection of light around it may provide the useful information about physical nature of the particular compact object and other related aspects. In this paper, we study the deflection of light by a dual charged stringy black hole spacetime derived in dilaton-Maxwell gravity. The variation of deflection angle with the impact parameter for different values of electric and magnetic charges is studied. We also study the shadow of this black hole spacetime to obtain the radius of shadow cast by it. We have considered an optically thin emission disk around it and  observed that there are not significant changes in the shadow cast by this black hole compared to well-known Schwarzschild black hole spacetime in GR. 

\end{abstract}


\maketitle

\section{Introduction}
General Relativity (GR) is a geometric theory of gravitation proposed by Einstein and has truly passed various experimental tests in the weak-field regime which makes GR a standard theory of gravitation in our universe \cite{Hartle2003,Joshi1993}. Black holes (BHs) which emerge as the solutions of Einstein's Field Equations (EFEs) are one of the peculiar compact objects predicted by GR in usual four dimensions (4D) \cite{Chandrasekhar1998}. Among one of the fascinating consequences of GR, the fields which have been emerged out to be exciting discoveries in astrophysics are gravitational lensing (GL) and black hole shadow. Recently, the Event Horizon Telescope(EHT) provided the first image of a BH in the center of galaxy M87 \cite{Akiyama:2019cqa,Akiyama:2019brx,Akiyama:2019sww,Akiyama:2019bqs,Akiyama:2019fyp,Akiyama:2019eap} which shows a bright ring with a dark, central spot. The ring is actually a bright disk of a gas orbiting in the galaxy M87 and the spot is the BH shadow. This study therefore supports the earlier mathematical studies for existence of BHs in our universe as well as provides a possible way to verify one of the fundamental predictions of GR by the virtue of GL. In fact, the shadow is defined as the region of the observer’s sky which is left dark if there are light sources distributed everywhere, including from behind the observer \cite{Gralla:2019xty}.\\

The GL is indeed one of the most interesting astronomical phenomena which was first tested and confirmed by Arthur Eddington in 1919 during a solar eclipse. The GL is an important tool to characterize various properties of BHs in GR and other alternative theories of gravity. Various aspects of GL for BH spacetimes are discussed by different groups in diverse contexts \cite{Liebes:1964zz,Refsdal:1993kf,Nzioki:2010nj,kuniyal2018strong,Uniyal:2018ngj,Azreg-Ainou:2017obt}. Bozza et al. first investigated analytically the GL in the Schwarzschild BH spacetime background in strong field limit \cite{Bozza:2001xd}. The technique has subsequently been applied time and again for different BHs and later generalized to an arbitrary static, spherically symmetric metric \cite{Bozza:2002zj}. Further, the theoretical aspects related to the contour of BH shadows also have been investigated since the null geodesics near BHs and other compact objects would determine several important properties of given spacetimes. The shape of a BH shadow for a non-rotating BH is generally a circular disk and there is no deformation  observed in shadow size. The angular diameter of the shadow for a Schwarzschild BH was calculated by Synge \cite{Synge:1966okc} as a function of the mass of the BH and of the radius coordinate where the observer is situated. The shape of BH shadow for rotating BH is however no longer circular and it varies with the rotation parameter of BH. The shape and size of the shadow of rotating BH was first calculated by Bardeen \cite{hawking1973black}. The analytical study of the shape and size of BH shadow for the whole class of Plebanski-Demianski spacetimes \cite{Grenzebach:2014fha} is successfully performed by by Perlic et.al \cite{Grenzebach:2015oea}. Semi-analytic calculations \cite{Johannsen:2015qca} and a new numerical method \cite{Younsi:2016azx} of the contour of the shadow of rotating BHs were investigated. Hioki and Maeda \cite{Hioki:2009na} also discussed a relation between the shape of contour of the shadow and the inclination angle and the spin parameter of the Kerr BH. Various other attempts to study the shadow in various theories of GR and other alternative theories of gravity have also made in recent times \cite{Narayan:2019imo,Schneider:2018hge,Cunha:2016wzk,Vagnozzi:2020quf,Tsupko:2019pzg,Papnoi:2014aaa,Atamurotov:2015nra}. The general formula for the radius of the shadow of spherically symmetric BHs was considered by Perlic et al. as well as more recently by Konoplya \cite{Perlick:2015vta,Konoplya:2019sns}. Some recent studies related to the shadows of the BHs are also focused towards the accretion disk properties and the images of the shadow cast by these radiated emissions \cite{10.1093/mnras/sty2624,Jaroszynski:1997bw}. Simulations of those accretions has thrown quite a bit light on the central host of the source.\\

The non-trivial geometry of different types of BHs is of crucial interest in GR and other alternative theories of gravity \cite{PhysRevD.99.104018}. All the four dimensional geometries, we choose to work with generally have a Schwarzschild limit (obtainable by setting the relevant parameter zero in the line element) and one such BH arise in context of string theory \cite{Garfinkle:1990qj,Horowitz:1992jp}. The formulation as mentioned in \cite{Perlick:2015vta,Konoplya:2019sns} can  therefore be applied to a dual charged stringy BH spacetime to observe the difference of the shape and size of the shadow cast by it along with its comparison with most general BH spacetime i.e. Schwarzschild BH in GR. It may further describe new examples of the contours of the shadows of such BHs in view of optically thin emission disk in a much finer way.\\

The outline of our paper is as follows. In section 2, we introduce the spacetime metric and obtained the equations of motion for light rays. Section 3 includes the bending angle and circular light orbits which are of crucial relevance for the formation of the shadow. In section 4, we have calculated the radius of photon sphere and then obtained the angular radius of the shadow. In section 5, we consider a simple model of accretion flow by considering an optically thin emission region surrounding the dual charge stringy BH to see the images of the shadows cast by such accretion flow. Finally,   conclusions draw are summarised in section 6.

\section{The Stringy BH Spacetimes and Photon Motion}
The well-known asymptotically flat solutions which represent BHs with the electric and dual magnetic charge in dilaton-Maxwell gravity are due to Garfinkle, Horowitz and Strominger (GHS) \cite{Garfinkle:1990qj,Horowitz:1992jp}.The 4D effective action to describe the above solutions reads as, 
\begin{equation}
S = \int d^{4}x \sqrt{-g} \left[-R + 2 (\nabla \phi)^{2} + e^{-2\phi} F^{2}\right],
\end{equation}
here, $ F_{\mu \nu} $ is Maxwell field associated with $ U(1) $ sub group of $ E_{8} \times E_{8}  $ and $ \phi $ is dilaton, it is a scalar field which couples to Maxwell field. 
The field equations yielded by extremizing the action with respect to $ U(1) $ potential $ A_{\mu} $, $ \phi $ and  $ g_{\mu \nu} $ are
\begin{equation}
	\nabla_{\mu} (e^{-2\phi} F^{\mu \nu}) = 0 \label{field equation 1},
\end{equation}   
\begin{equation}
	\nabla^{2} \phi + \frac{1}{2}e^{-2\phi} F^{2} = 0, \label{field equation 2}
\end{equation} 
\begin{equation}
	R_{\mu \nu} = 2 \nabla_{\mu} \phi \nabla_{\nu} \phi + 2 e^{-2\phi} F_{\mu \rho} F_{\nu}^{\rho} - \frac{1}{2} g_{\mu \nu} e^{-2\phi} F^{2}. \label{field equation 3}
\end{equation}
For $ F_{\mu\nu} = 0$, it reduces to standard Einstein scalar field action as in GR. 
 The spacetime geometry arises from the field equations (\ref{field equation 1} - \ref{field equation 3})  have  line elements those are causally similar to Schwarzschild geometry in GR as evident from the metric given below. The metric for the BH with electric charge is given as \cite{Dasgupta:2008in},
\begin{equation}
ds_{E}^2 = - \frac{\left(1-\frac{m}{r}\right)}{\left(1+ \frac{m \sinh^2{\alpha}}{r}\right)}dt^2 + \frac{dr^2}{\left(1-\frac{m}{r}\right)} + r^2 {d\Omega}_2^2, \label{metric1}
\end{equation}	
where ${d\Omega}_2^2 = (d\theta^2 + \sin{\theta}^2 d\phi^2)$ is the metric on a two dimensional unit sphere and $\alpha$ is a parameter related to the electric charge. The magnetically charged BH can be obtained from the electrically charged solution by electromagnetic duality transformation \cite {Horowitz:1992jp} and the metric  with a magnetic charge parameter is given as follows,
\begin{eqnarray}
ds_{M}^2 = - \frac{\left(1-\frac{m}{r}\right)}{\left(1- \frac{Q^2}{mr}\right)}dt^2 + \frac{dr^2}{\left(1-\frac{m}{r}\right)\left(1-\frac{Q^2}{m r}\right)} + r^2 d{\Omega}_2^2.  \label{metric2} 
\end{eqnarray}
Here $Q$ is the magnetic charge of the BH. In the respective limits i.e. $\alpha=0$ and $Q = 0$, the Schwarzschild BH geometry as in GR in both the cases can be reproduced easily. 

Since the metric is spherically symmetric so each plane can be considered  as an equatorial one, so that one can choose $\theta = \frac{\pi}{2}$ and consequently $\dot\theta = 0$. 
Using the Hamilton-Jacobi equation, the equation of motion can be derived easily. The Hamiltonian for light rays given as,
\begin{equation}
H = \frac{1}{2} g^{ik} p_{i}p_{k} = \frac{1}{2} \left(-\frac{p_{t}^2}{g_{tt}} + \frac{p_{r}^2}{g_{rr}} + \frac{p_{\phi}^2}{g_{\phi\phi}}\right). \label{hamiltonian}
\end{equation} 
The light rays are the solutions to the equation of motion with,
\begin{equation}
\dot{p_i} = -\frac{\partial H}{\partial x^i} ,  \dot{x^i} = \frac{\partial H}{\partial p_{i}}.
\end{equation}
We will now discuss the photon momentum and effective potentials for the cases corresponding to the line elements (\ref{metric1}) and (\ref{metric2}) respectively.\\

\noindent\textbf{Case I} [Corresponding to line element (\ref{metric1})]-\\

\noindent The momenta for electric charge stringy BH given as follows, 
\begin{equation}
\dot{t} = -\frac{p_t}{g_{tt}} = -E \left(1+ \frac{m \sinh^2\alpha}{ r}\right)\left(1-\frac{m}{r}\right)^{-1},
\end{equation}
\begin{equation}
\dot{\phi} = -\frac{p_\phi}{g_{\phi\phi}} = \frac{L}{r^2},
\end{equation}
\begin{equation}
\dot{r} = \frac{p_r}{g_{rr}} = p_{r} \left(1-\frac{m}{r}\right). \label{r1}
\end{equation}
Here $E$ and $L$ are represent the energy and angular momentum of test particle respectively.
The effective potential for this case reads accordingly as below, 
\begin{equation}
V_{E}(r) = \frac{L^2}{r^2}\frac{\left( 1- \frac{m}{r} \right)}{\left( 1+ \frac{m \sinh^2\alpha}{r} \right)}. \label{ve}
\end{equation}
\textbf{Case II} [Corresponding to line element (\ref{metric2})]-\\

\begin{figure*}
	\centering
	\includegraphics[scale=0.48]{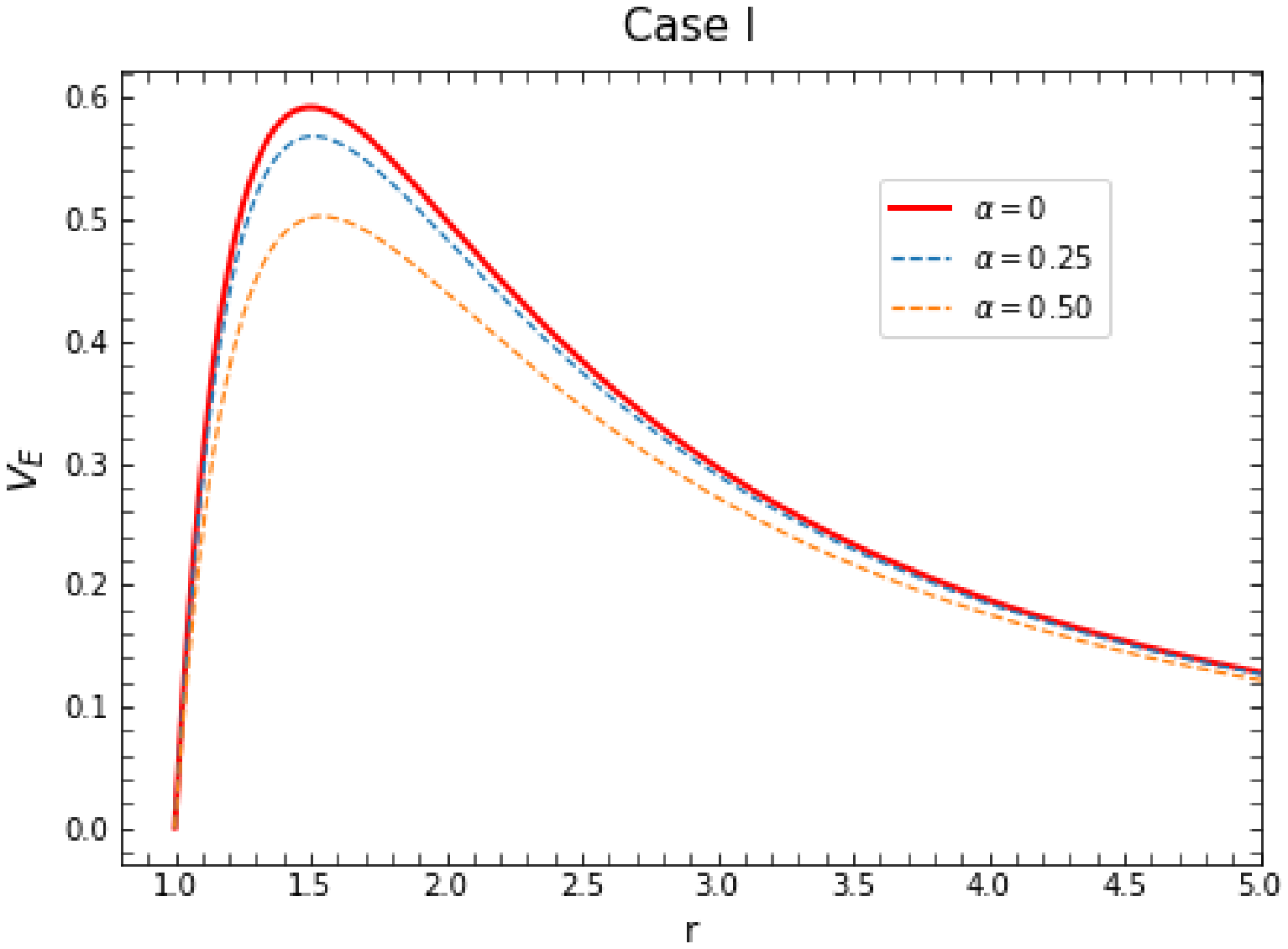}
	\includegraphics[scale=0.48]{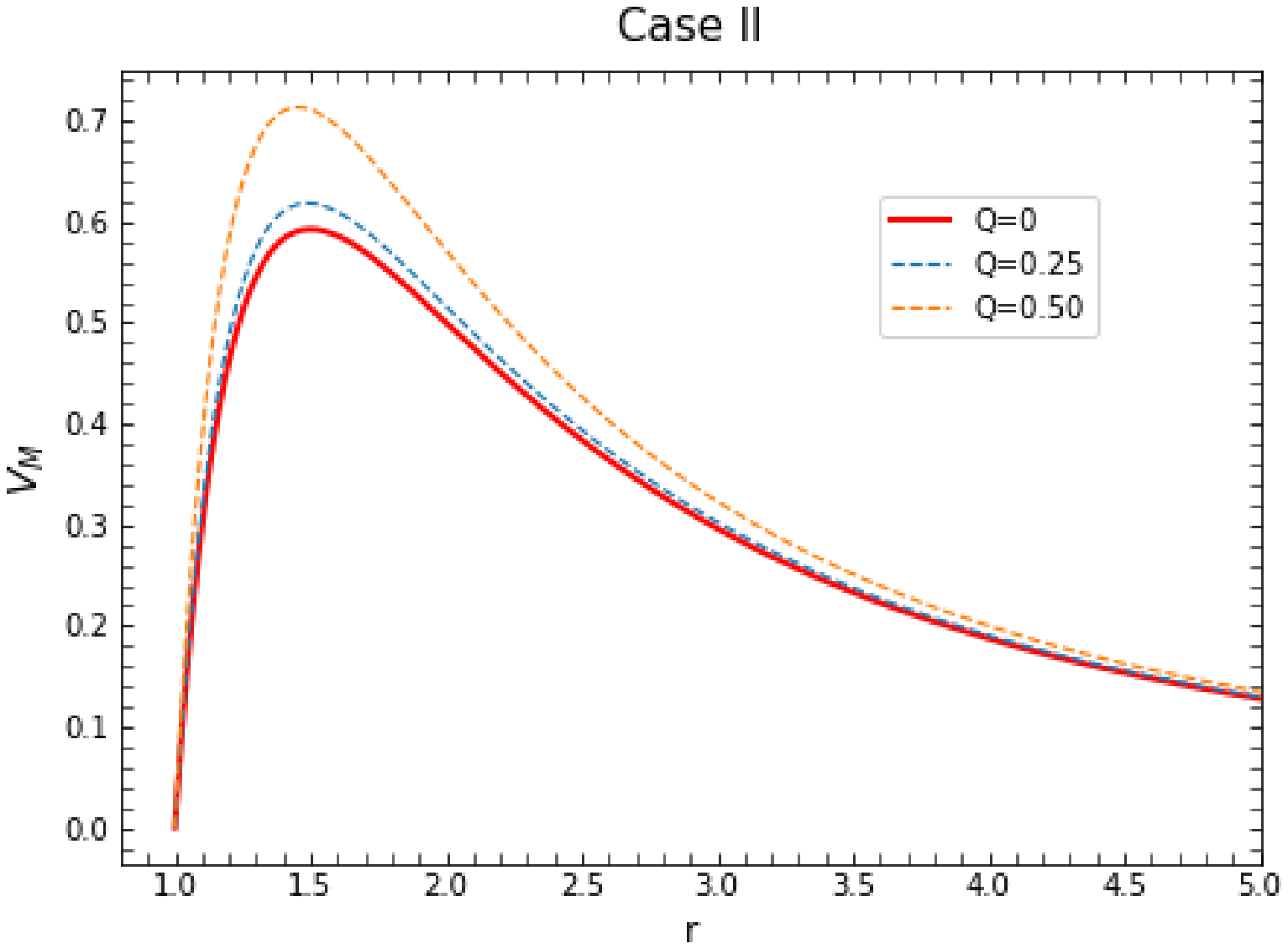}
	\caption{Effective potentials of the Stringy BH for different values of electric (Case I) and magnetic charge (Case II) respectively. The cases with $\alpha=0$, $Q=0$ correspond to the Schwarzschild BH geometry in GR.}
	\label{fig:my_label}
\end{figure*}

\noindent The momenta for magnetic charge stringy BH is given as follows,
\begin{equation}
\dot{t} = -\frac{p_t}{g_{tt}} = -E \left(1- \frac{Q^2}{m r}\right)\left(1-\frac{m}{r}\right)^{-1},
\end{equation}
\begin{equation}
\dot{\phi} = -\frac{p_\phi}{g_{\phi\phi}} = \frac{L}{r^2}, \label{phi}
\end{equation}
\begin{equation}
\dot{r} = \frac{p_r}{g_{rr}} = p_{r} \left(1-\frac{m}{r}\right)\left(1- \frac{Q^2}{m r}\right). \label{r} 
\end{equation}
The effective potential for this case reads as, 
\begin{equation}
V_{M}(r) = \frac{L^2}{r^2}\frac{\left( 1- \frac{m}{r} \right)}{\left( 1- \frac{Q^2}{m r} \right)}. \label{vm}
\end{equation}
For $H=0$, the Hamiltonian for light rays as described by equation (\ref{hamiltonian}) leads to,
\begin{equation}
\left(-\frac{p_{t}^2}{g_{tt}} + \frac{p_{r}^2}{g_{rr}} + \frac{p_{\phi}^2}{g_{\phi\phi}}\right) = 0. \label{hzero}
\end{equation}
such that the radial momentum is given as follows,
\begin{equation}
p_{r}^2 =  p_{t}^2 \frac{g_{rr}}{g_{tt}} - p_{\phi}^2 \frac{g_{rr}}{g_{\phi\phi}}. \label{pr}  
\end{equation}
Here the dot represents the derivatives with respect to an affine parameter and a prime denotes the derivatives with respect to r. The momenta $p_t$ and $p_\phi$ are constants of motion and thus represent the energy and angular momentum of light rays respectively. From equations (\ref{r}) and (\ref{phi}), one can obtain,
\begin{equation}
\frac{dr}{d \phi} = \frac{\dot{r}}{\dot{\phi}} = \frac{(g_{\phi\phi}) p_r}{(g_{rr}) L },
\end{equation}
and using $p_{r}$ from (\ref{pr}), 
\begin{equation}
\frac{d r}{d \phi} = \pm \frac{\sqrt{g_{\phi\phi}}}{\sqrt{g_{rr}}} \sqrt{\frac{E^2}{L^2} h(r)^2 -1} \label{drdphi}
\end{equation}
where, the function $h(r)$ for electric and magnetic charge BH is calculated respectively as given below,
\begin{equation}
h_{E}(r)^2 = \frac{g_{\phi\phi}}{g_{tt}} = \frac{r^2\left(1+\frac{m \sinh^2\alpha}{ r}\right)}{\left(1-\frac{m}{r}\right)},  
\end{equation}
and
\begin{equation}
h_{M}(r)^2 = \frac{g_{\phi\phi}}{g_{tt}} = \frac{r^2\left(1-\frac{Q^2}{m r}\right)}{\left(1-\frac{m}{r}\right)}.
\end{equation}\\
The effective potentials as given in equations (\ref{ve}) and (\ref{vm}) for Case I and Case II have been already discussed in \cite{Dasgupta:2008in} in detail. In order to observe the effect of dual charge, we present effective potential of the same BH for different values of electric charge($\alpha$) and magnetic charge($Q$). The effective potentials for different values of $\alpha$ and $Q$ and their comparison with Schwarzschild BH is shown in Fig.\ref{fig:my_label}. It can easily observed from effective potential that as the value of electric charge parameter increases the value of effective potential decreases (see Fig.\ref{fig:my_label} Case I) and vice-versa for magnetic charge parameter (see Fig.\ref{fig:my_label} Case II ). There is no minima present in each curve therefore there exist no stable orbit for the photons and only unstable orbits may exist in each case.
\section{Bending Angle and Circular Light Orbits}
Let us consider a light ray that comes in from infinity, reaches a minimum at a radius R, and goes out to infinity again. With such consideration, the integration over the orbit then leads to the following formula \cite{Perlick:2015vta} for the bending angle $\delta$,
\begin{equation}
\delta = -\pi + 2\int_{R}^{\infty} \frac{\sqrt{g_{rr}}}{\sqrt{g_{\phi\phi}}} \frac{dr}{\sqrt{\left(\frac{p_{t}^2}{p_{\phi}^2} h(r)^2 -1\right)}}.
\end{equation}
Here R corresponds to the turning point of the trajectory and
the condition $\frac{d r}{d \phi}\vert_{R=0}$ needs to hold necessarily. This equation relates $R$ to the constant of motion $p_{\phi}/p_{t}$ as below,
\begin{equation}
h_{M}(R)^2 = h_{E}(R)^2 =  \frac{p_{\phi}^2}{p_{t}^2} = \frac{L^2}{E^2}.
\end{equation}
The deflection angle may then be rewritten as a function of $R$,
\begin{equation}
\delta = -\pi + 2\int_{R}^{\infty} \frac{\sqrt{g_{rr}}}{\sqrt{g_{\phi\phi}}} \frac{dr}{\sqrt{\left(\frac{h(r)^2}{h(R)^2}  -1\right)}}. \label{delta}
\end{equation}
The expression of bending angle given by equation (\ref{delta}) takes the form for the \textbf{Case I} as,
\begin{equation}
\delta_{E} = -\pi + 2\int_{R}^{\infty}  \frac{dr}{\sqrt{r(r-m)} \sqrt{\left( \frac{r^2 \left(1+\frac{m \sinh^2{\alpha}}{r}\right)}{b^2(1-m/r)} -1 \right)}}.\label{deltae}
\end{equation}

\begin{figure*}
	\centering
	\includegraphics[scale=0.45]{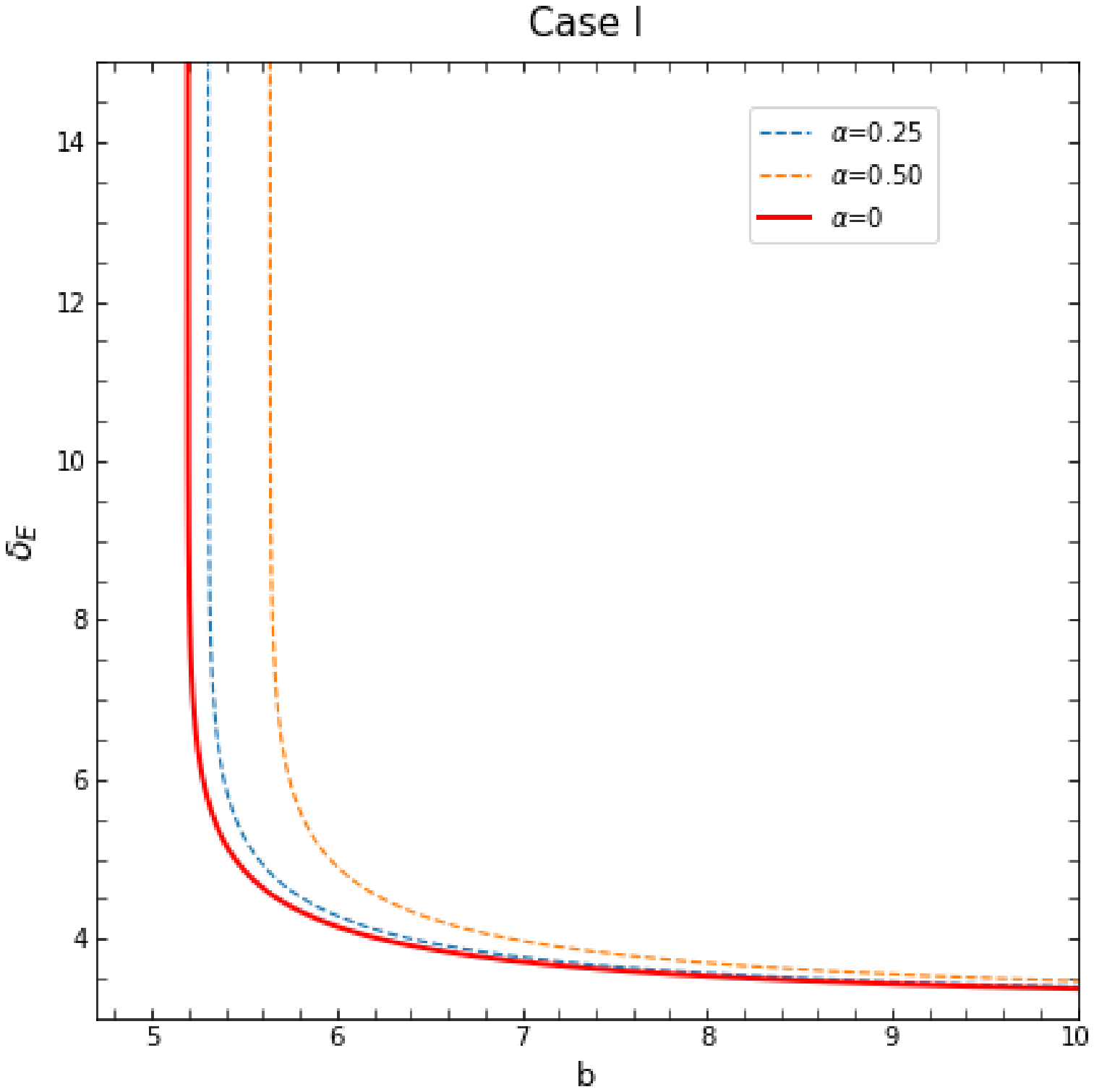}
	\includegraphics[scale=0.45]{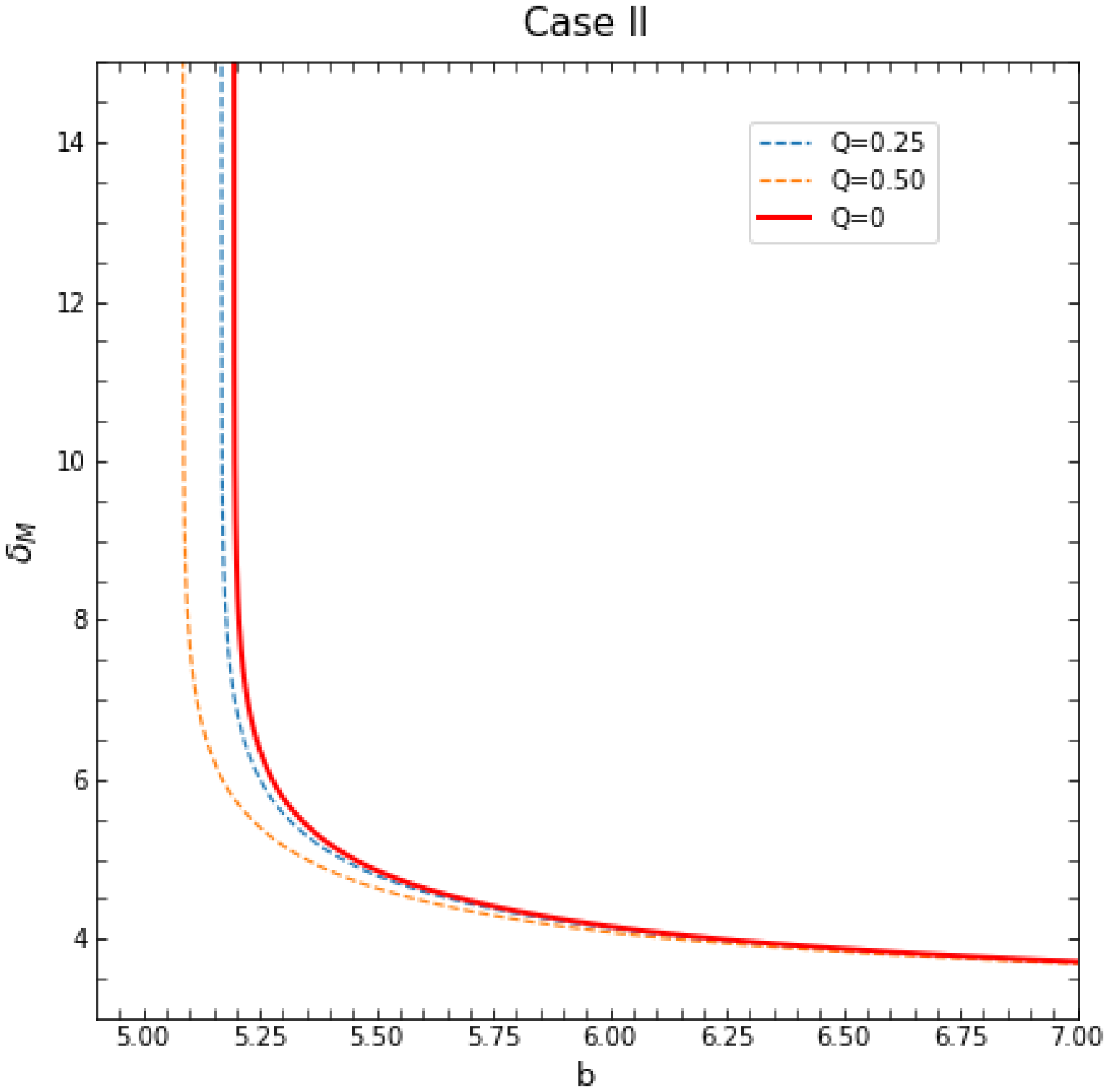}
	\caption{Deflection angle with the impact parameter of stringy BH for different values of $\alpha$ (Case I) and Q (Case II) with $m=1$. Here, the impact parameter is a function of the closest distance of approach.}
	\label{FIG:2}
\end{figure*}

However, the deflection angle for \textbf{Case II} reads as,
\begin{equation}
\delta_{M} = -\pi + 2\int_{R}^{\infty} \frac{dr}{\sqrt{r (r-m)(1-\frac{Q^2}{m r})} \sqrt{\left(\frac{r^2 \left(1-Q^2/m r\right)}{b^2 (1-m/r)} -1 \right)}}.\label{deltam}
\end{equation}
where $b=L/E$ is the impact parameter.
The expression obtained in equation (\ref{deltae}) and (\ref{deltam}) with prescribed limits reduce to Schwarzschild BH case in GR. The plots represent the variation of deflection angle with impact parameter of BH for magnetic and electric charged stringy BH (see Fig.\ref{FIG:2}). The cases with vanishing of charge parameters correspond to the Schwarzschild case as studied in \cite{Gerard:1999wd}. In Fig.(\ref{FIG:2}), every single curve for both cases indicate that, by increasing the value of impact parameter, the bending angle decreases for different values of $Q$ and $\alpha$. However, with an increase in the value of charge parameters, one can observe that the value of closest approach show the lateral inversion behaviour with one another. For magnetic charge ($Q$), the critical value of the closest approach decreases since the light goes closer to the BH however vice-versa for electric charge ($\alpha$).\\
A circular light orbits corresponds to zero radial velocity and acceleration, so that $\dot{r} = 0$ and $\ddot{r} = 0$. We will now discuss the circular light orbits for the above mentioned cases.\\
\noindent\textbf{Case I}-\\

\noindent Using $p_{r} = 0$ in equation (\ref{hzero}) along with equation (\ref{r1}), we obtain
\begin{equation}
\frac{E^2 \left( 1+ \frac{m \sinh^2\alpha}{r} \right)}{(1-\frac{m}{r})} = \frac{L^2}{r^2},
\end{equation}
and the angular momentum term can be separated as follows,
\begin{equation}
L_{E}^2 = \frac{E^2r^2(1 + \frac{m \sinh^2 {\alpha}}{r})}{(1- m/r)}.\label{le1} 
\end{equation}
Differentiating equation (\ref{hzero}) w.r.t affine parameter along with ${p_{r}} = 0$, we have
\begin{equation}
\frac{m E^2(1+ \sinh^2\alpha)r^3 - 2L^2(m-r)^2}{(m-2)^2 r^3} =0.
\end{equation}
and separating the angular momentum term from above expression, as 
\begin{equation}
L_{E}^2 =  \frac{m E^2(1 + \sinh^2 {\alpha})r^3}{(m-r)^2}. \label{le2}
\end{equation}

\noindent\textbf{Case II}-\\

\noindent Using $p_{r} = 0$ in equation (\ref{hzero}) along with equation (\ref{r}), we obtain
\begin{equation}
E^2 \left(1- \frac{Q^2}{m r}\right)\left(1-\frac{m}{r}\right)^{-1} = \frac{L^2}{r^2}, 
\end{equation}
and separating the angular term leads to,
\begin{equation}
L_{M}^2 = \frac{r^2 E^2 (1-Q^2/m r)}{(1-m/r)}. \label{lm1}
\end{equation}
Differentiating equation \ref{hzero} w.r.t affine parameter along with ${p_{r}} = 0$, we have
\begin{equation}
\frac{-2 m L^2 (m - r)^2 + (m^2 - Q^2) E^2 r^3}{m (m - r)^2 r^3} = 0,
\end{equation}
and once again separating the angular momentum term, we have
\begin{equation}
L_{M}^2 = \frac{r^3 E^2(m^2 - Q^ 2)}{2 m (m-r)^2}.\label{lm2}
\end{equation}
Subtracting the set of these two equations i.e. (\ref{le1}), (\ref{le2}) and (\ref{lm1}),  (\ref{lm2}) from each other yields, the radius of a circular light orbit in the form
\begin{equation}
0 = \frac{d}{d r} h(r)^2. \label{dhdr}
\end{equation}
The solution of equation (\ref{dhdr}) at $r = r_{ph}$ gives the radius of photon sphere for both the cases as discussed in detail in the forthcoming section.
\section{Radius of the Shadow}
The motion of light rays near the photon sphere are crucial to understand the properties of BH shadows. The radius $r_{ph}$ of the outermost photon sphere is the critical $r_{ph}$ value of the minimal radius R mentioned above. If a light ray comes in from infinity and reaches a minimum radius R bigger than $r_{ph}$, it will go out to infinity again. The case $R = r_{ph}$ corresponds to a light ray that spirals asymptotically towards a circular photon orbit in the sphere of radius $r_{ph}$. We consider a light ray that is sent from the observer’s position at $r_{0}$ into the past under an angle $\beta$ \cite{Perlick:2015vta} with respect to the radial direction such that
\begin{equation}
\cot{\beta} = \frac{\sqrt{g_{rr}}}{\sqrt{g_{\phi\phi}}} \frac{d r}{d \phi}\vert_{R=0}. \label{cot}
\end{equation}
The equation (\ref{drdphi}) may then be written in terms of minimal radius $R$ as follows 
\begin{equation}
\frac{d r}{d \phi} = \pm \frac{\sqrt{g_{rr}}}{\sqrt{g_{\phi\phi}}} \sqrt{\frac{h^2(r)}{h^2(R)} - 1}.  \label{drdphi2}
\end{equation}
Using equations (\ref{cot}) and (\ref{drdphi2}),
\begin{equation}
\sin^2{\beta} = \frac{h(R)^2}{h(r_{0})^2}.\label{angular} 
\end{equation} 
The boundary of the shadow $\beta_{s}$ is determined by light rays that spiral asymptotically towards a circular light orbit at radius $r_{ph}$. We will now first discuss the photon orbits for the Case I and Case II to obtain $\beta_{s}$. \\

\noindent\textbf{Case I}-\\

\noindent From equation (\ref{dhdr}), the radius of photon sphere for BH with electric charge is calculated as,
\begin{equation}
r_{ph}^E = \frac{1}{4} \left[ 3m - m\sinh^2 {\alpha} + \frac{ m\cosh{\alpha}  \sqrt{17 + \cosh{(2 \alpha)}}}{\sqrt{2}} \right]. \label{rphe}
\end{equation}
The critical value of minimal radius $R$ is however given as,
\begin{equation}
R_{E} = \frac{r^E_{ph}(r^E_{ph} + m \sinh^2\alpha)}{\left(1-\frac{m}{r^{E}_{ph}}\right)}.
\end{equation}
The angular radius of the shadow of BH with electric charge as described by equation (\ref{angular}) is then obtained as follows,
\begin{equation}
\sin^2{\beta_{s}^E} = \frac{h(r_{ph})^2}{h(r_{0})^2} = \frac{r_{ph}^2 \left(1+\frac{m \sinh^2{\alpha}}{r_{ph}}\right) \left(1-\frac{m}{r_{0}}\right)}{ r_{0}^2\left(1-\frac{m}{r_{ph}}\right) \left(1+\frac{m \sinh^2{\alpha}}{r_{0}}\right)}, \label{sine}
\end{equation}
where $r_{ph}$ has to be determined from equation (\ref{rphe}).\\

\noindent\textbf{Case II}-\\

\noindent The radius of photon sphere for BH with magnetic charge is calculated as,
\begin{equation}
r_{ph}^M = \frac{(3m/2 + Q^2/2m) + \sqrt{9m^2/4 + Q^4/4m^2 + 3Q^2/2  - 4Q^2}}{2}. \label{rphm}
\end{equation}
The critical value of the minimal radius $R$ thus reads as,
\begin{equation}
R_{M} = \frac{r^M_{ph} \left( r^M_{ph} - \frac{Q^2}{m} \right)}{\left( 1- \frac{m}{r^M_{ph}} \right)}.
\end{equation}
Therefore the angular radius of the shadow of BH with magnetic charge is then obtained as follows,
\begin{equation}
\sin^2{\beta_{s}^M} = \frac{h(r_{ph})^2}{h(r_{0})^2} = \frac{r_{ph}^2 \left( 1-\frac{m}{r_{0}} \right) \left( 1-\frac{Q^2}{m r_{ph}} \right)}{r_{0}^2 \left( 1-\frac{m}{r_{ph}} \right) \left( 1 - \frac{Q^2}{m r_{0}} \right)}, \label{sinm}
\end{equation}
where $r_{ph}$ has to be determined from equation (\ref{rphm}).\\

The expressions obtained in equations (\ref{sine}) and (\ref{sinm}) with prescribed limits reduce to Schwarzschild BH case in GR \cite{Synge:1966okc}.

\section{Optically thin emission disk surrounded by a Stringy BH}

\begin{figure*}
	\centering
	\subfigure[]{\includegraphics[scale=0.47]{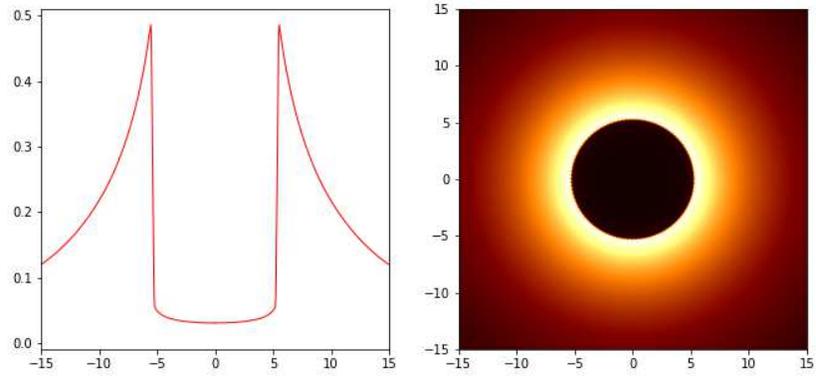}\label{fig:image1}}
	\subfigure[]{\includegraphics[scale=0.47]{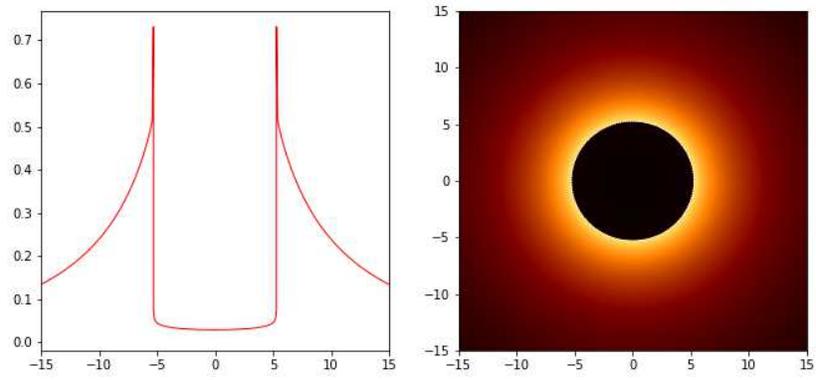}\label{fig:image2}}
	\subfigure[]{\includegraphics[scale=0.47]{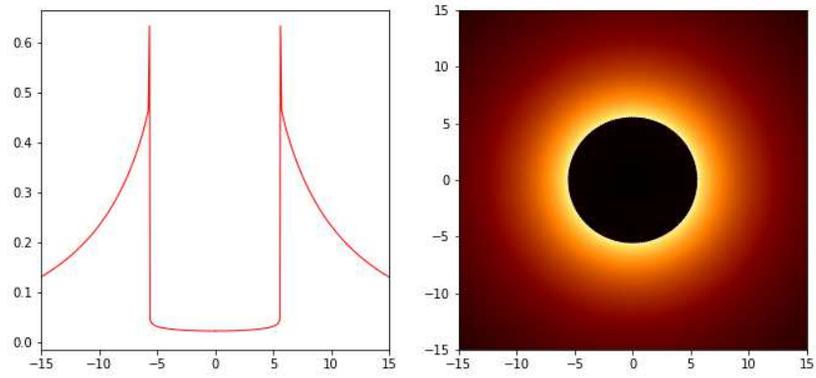}\label{fig:image3}}
	\caption{(Case I)-The first column represents the intensity distribution with respect to the impact parameter and the second column represents images of optically thin emission region surrounding the electrically charge stringy BH with $m=1$, for (a) $\alpha=0$ (i.e. Schwarzschild BH), (b) $\alpha=0.25$ and (c) $\alpha=0.50$.}
	\label{fig:images}
\end{figure*}

\begin{figure*}
	\centering
	\subfigure[]{\includegraphics[scale=0.47]{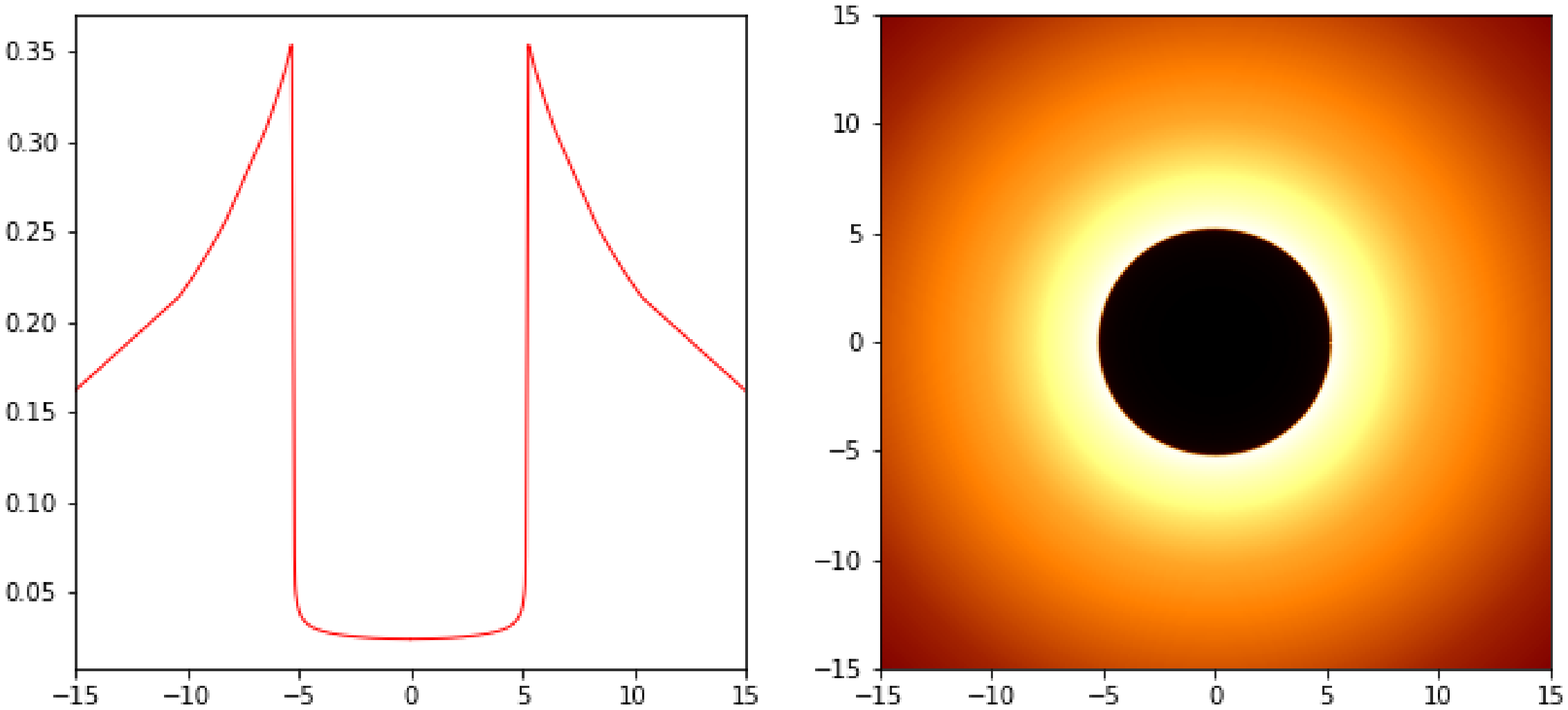}\label{fig:image4}}	\subfigure[]{\includegraphics[scale=0.47]{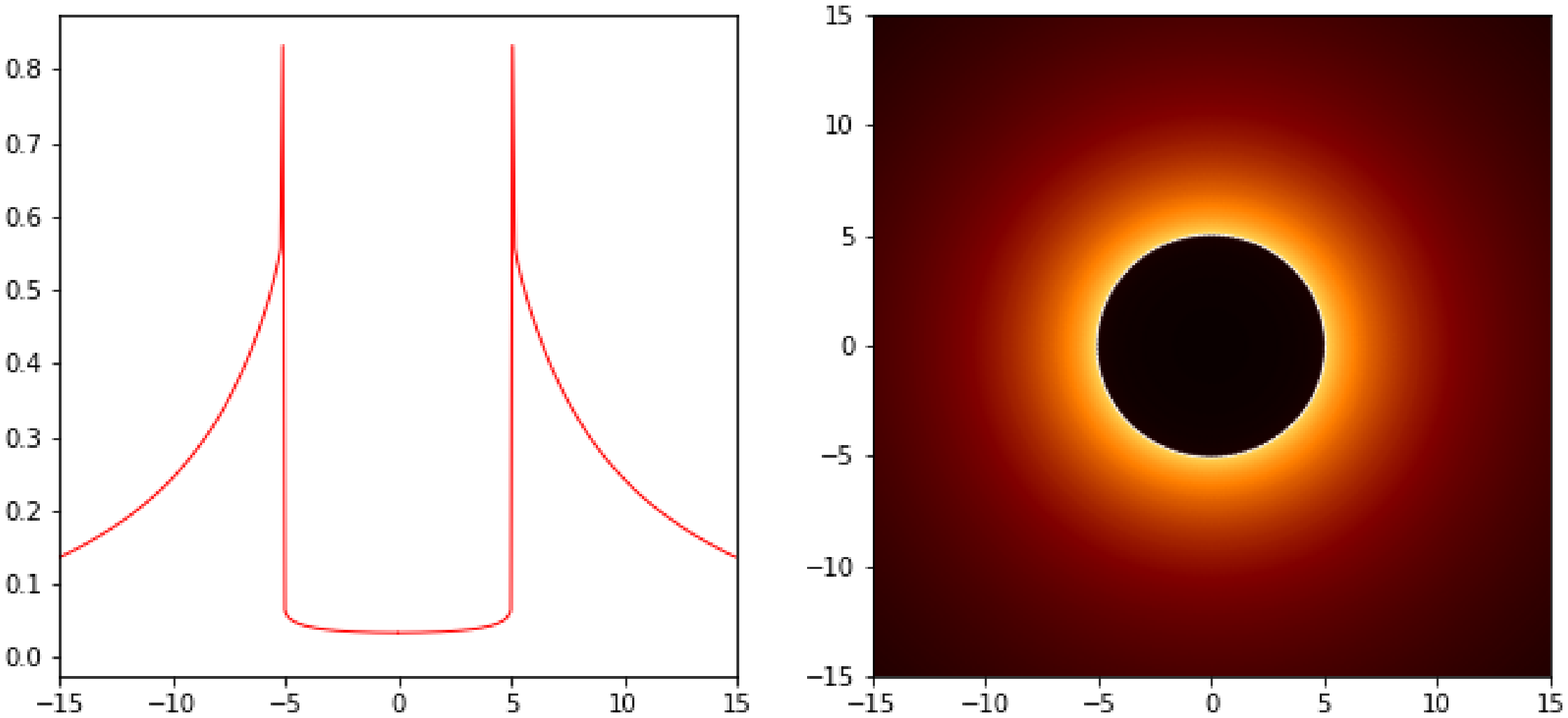}\label{fig:image5}}
	\caption{(Case-II)-The first column represents the intensity distribution with respect to the impact parameter and the second column represents images of optically thin emission region surrounding the magnetically charge stringy BH with $m=1$, for (a) $Q=0.25$ and (b) $Q=0.50$. The Schwarzschild BH case is similar as in Fig.\ref{fig:image1}(Case I).}
\end{figure*}

\noindent Here, we consider a simple model of optically thin, radiating accretion flow around the stringy BH. For the emission mechanisms, certain assumptions shall be made for the calculation of the intensity from the radiating accretion flow.
The observed specific intensity $I_{\nu 0}$ at the observed photon frequency $\nu_\text{obs}$ at the point $(X,Y)$ of the observer's image (usually measured in $\text{erg} \text{s}^{-1} \text{cm}^{-2} \text{str}^{-1} \text{Hz}^{-1}$) is given by\cite{Jaroszynski:1997bw,PhysRevD.87.107501},

\begin{eqnarray}
I_{obs}(\nu_{obs},X,Y) = \int_{\gamma}g^3 j(\nu_{e})dl_\text{prop},  \label{inte}
\end{eqnarray}
where $g = \nu_{obs}/\nu_{e}$ is the redshift factor, $\nu_{e}$ is the photon frequency as measured in the rest-frame of the emitter, $j(\nu_{e})$ is the emissivity per unit volume in the rest-frame of the emitter, and $dl_\text{prop} = k_{\rho}u^{\rho}_{e} d\lambda$ is the infinitesimal proper length as measured in the rest-frame of the emitter . The redshift factor as in equation (\ref{inte}) is calculated from,
\begin{equation}
g = \frac{k_{\rho}u^{\rho}_{\text{obs}}}{k_{\sigma}u^{\sigma}_{e}}, \label{rdshift}
\end{equation}
where $k^{\mu}$ is the 4-velocity of the photons, $u^{\rho}_{e}$ 4-velocity of the accreting gas emitting the radiation and  $u^{\mu}_{\text{obs}}$ is 4-velocity of the observer with $\lambda$ is the affine parameter along the photon path $\gamma$. The whole integral given by equation (\ref{inte}) should be evaluated over the path $\gamma$ of the photons i.e. (null geodesics). In order to generalise the formalism for both the cases, we will restrict to the equatorial plane only and assume that the gas is radially free falling with a four-velocity whose components are described below,
\begin{eqnarray}
u^t_{e} & = & \frac{1}{g_{tt}(r)}, \nonumber \\
u^r_{e} & = & -\sqrt{\frac{1-g_{tt}(r)}{g_{tt}(r)g_{rr}(r)}}, \nonumber \\
u^{\theta}_{e} & = & 0, \nonumber \\
u^{\phi}_{e} & = & 0, 
\end{eqnarray}
where the metric components $g_\textit{tt}$, $g_\textit{rr}$ and $g_{\phi\phi}$ are corresponding to the line elements for Case I and Case II. 
In previous section, the components of four-velocity were already calculated. In order to ease further calculations, we obtain an equation between the radial and time component of the four-velocity as follows,
\begin{equation}
\frac{k_r}{k_t} = \pm \sqrt{g_{rr}\bigg(\frac{1}{g_{tt}}-\frac{b^2}{g_{\phi\phi}}\bigg)},
\end{equation}
where the sign $ +(-) $ corresponds to photon approaches (goes
away) from the massive object.
The redshift function $g$ is then given by,
\begin{eqnarray}
g =\frac{1}{\frac{1}{g_{tt}} \pm \frac{k_r}{k_t}\sqrt{\bigg(\frac{1-g_{tt}}{g_{tt}g_{rr}}\bigg)}}.
\end{eqnarray}

The specific emissivity for a simple model in which the emission is monochromatic with emitter's-rest frame frequency $\nu_{\star}$, and the emission has a $1/r^2$ radial profile is calculated as,
\begin{equation}
j(\nu_{e}) \propto \frac{\delta_D(\nu_{e}-\nu_{\star})}{r^2},
\end{equation}
where $\delta_D$ is the Dirac delta function. The proper length can be written as, 
\begin{equation}
dl_{\text{prop}} = k_{\rho}u^{\rho}_{e}d\lambda = -\frac{k_t}{g|k^r|}dr,
\end{equation}
Integrating the Intensity over all the observed frequencies, we obtain the observed photon intensity as given below,
\begin{equation}
I_{obs}(X,Y) \propto -\int_{\gamma} \frac{g^3 k_t}{r^2k^r}dr.
\end{equation}
Now we have the desired intensity equation so we can proceed to compute the images. The numerical routine was implemented in a python package \texttt{EinsteinPY} \cite{shreyas_bapat_2019_3526198} and modified for the use of stringy BH spacetimes as Case I and Case II accordingly.

\section{Summary, Conclusion and Future Directions}
In this paper, we have studied deflection angle and BH shadow of a dual charged stringy BH. First, we have analysed the null geodesics and the motion of a photon around this BH spacetime in brief. Using Hamilton-Jacobi formalism, we find out the equations of motion and the deflection angle of light around the above stringy BH spacetimes accordingly. The maximum bending angle takes place at a critical value of impact parameter and this critical value increases with an increase in the electric charge. In contrast, in case of magnetically charge stringy BH, the critical value for maximum deflection angle shifts towards a lower value as magnetic charge increases. However, with an increase in the value of electric and magnetic charge parameters as compared to the Schwarzschild BH case, it is observed that the value of closest approach shows the lateral inversion behaviour with one another. Further, we obtained the circular orbits for photons which gives us the necessary condition that whether it forms an unstable photon sphere or not.  An unstable photon sphere constitutes for the shadow of the BH and the radius of the photon spheres is calculated accordingly for both the cases of dual charged stringy BH spacetime. The boundaries of the shadow were then found along with the angular radius. It is observed that the obtained expressions of radius of photon sphere and angular size of BH shadow with prescribed limits reduce to the Schwarzschild BH geometry in GR. Finally, we consider an optically thin emission disk around the BH (simple inverse square radial profile) and see the images produced by it. The images (see Figs.3, 4) produced in it show that there is not much significant changes for different values of electric and magnetic charge. The image while compared to the case of Schwarzschild BH seems indistinguishable however difference in deflection angle of a stringy BH for both cases as compared to Schwarzschild BH is clearly distinguishable. We therefore need a careful attention to distinguish among the geometries of stringy BHs and Schwarzschild BH from the view point of their optical and other properties.The stability of circular geodesics of these stringy charged BHs needs careful attention and we intended to report on this issue in near future. Further, there will also be detail investigation of lensing and shadow phenomenon for a Kerr-like BHs in various alternative theories of gravity in our future studies.

\section{Acknowledgments}
The authors are grateful to the anonymous referee for the valuable comments and suggestions which helped us to improve the quality of this paper.
The author SK is thankful to the Uttarakhand State Council of Science and Technology (UCOST), Dehradun for financial assistance through R\&D grant number UCS\&T/RD-18/18-19/16038/4. The author HN and PS acknowledge the financial support provided by Science and Engineering Research Board (SERB), New Delhi through the grant number EMR/2017/000339. The authors (SK, HN and PS) also acknowledge the facilities at ICARD, Gurukula Kangri Vishwavidyalaya, Haridwar. The authors also thank to Dr. Rajibul Shaikh for useful suggestions. 

\bibliographystyle{elsarticle-num}
\bibliography{mybibliography}

\begin{thebibliography}{10}
\expandafter\ifx\csname url\endcsname\relax
  \def\url#1{\texttt{#1}}\fi
\expandafter\ifx\csname urlprefix\endcsname\relax\def\urlprefix{URL }\fi
\expandafter\ifx\csname href\endcsname\relax
  \def\href#1#2{#2} \def\path#1{#1}\fi

\bibitem{Hartle2003}
J.~B. Hartle, Gravity: An introduction to einstein’s general relativity
  (2003).

\bibitem{Joshi1993}
P.~S. Joshi, Global aspects in gravitation and cosmology., Int. Ser. Monogr.
  Phys. 87.

\bibitem{Chandrasekhar1998}
S.~Chandrasekhar, The mathematical theory of black holes, Vol.~69, Oxford
  university press, 1998.

\bibitem{Akiyama:2019cqa}
K.~Akiyama, et~al., {First M87 Event Horizon Telescope Results. I. The Shadow
  of the Supermassive Black Hole}, Astrophys. J. 875~(1) (2019) L1.
\newblock \href {http://arxiv.org/abs/1906.11238} {\path{arXiv:1906.11238}},
  \href {https://doi.org/10.3847/2041-8213/ab0ec7}
  {\path{doi:10.3847/2041-8213/ab0ec7}}.

\bibitem{Akiyama:2019brx}
K.~Akiyama, et~al., {First M87 Event Horizon Telescope Results. II. Array and
  Instrumentation}, Astrophys. J. 875~(1) (2019) L2.
\newblock \href {http://arxiv.org/abs/1906.11239} {\path{arXiv:1906.11239}},
  \href {https://doi.org/10.3847/2041-8213/ab0c96}
  {\path{doi:10.3847/2041-8213/ab0c96}}.

\bibitem{Akiyama:2019sww}
K.~Akiyama, et~al., {First M87 Event Horizon Telescope Results. III. Data
  Processing and Calibration}, Astrophys. J. 875~(1) (2019) L3.
\newblock \href {http://arxiv.org/abs/1906.11240} {\path{arXiv:1906.11240}},
  \href {https://doi.org/10.3847/2041-8213/ab0c57}
  {\path{doi:10.3847/2041-8213/ab0c57}}.

\bibitem{Akiyama:2019bqs}
K.~Akiyama, et~al., {First M87 Event Horizon Telescope Results. IV. Imaging the
  Central Supermassive Black Hole}, Astrophys. J. 875~(1) (2019) L4.
\newblock \href {http://arxiv.org/abs/1906.11241} {\path{arXiv:1906.11241}},
  \href {https://doi.org/10.3847/2041-8213/ab0e85}
  {\path{doi:10.3847/2041-8213/ab0e85}}.

\bibitem{Akiyama:2019fyp}
K.~Akiyama, et~al., {First M87 Event Horizon Telescope Results. V. Physical
  Origin of the Asymmetric Ring}, Astrophys. J. 875~(1) (2019) L5.
\newblock \href {http://arxiv.org/abs/1906.11242} {\path{arXiv:1906.11242}},
  \href {https://doi.org/10.3847/2041-8213/ab0f43}
  {\path{doi:10.3847/2041-8213/ab0f43}}.

\bibitem{Akiyama:2019eap}
K.~Akiyama, et~al., {First M87 Event Horizon Telescope Results. VI. The Shadow
  and Mass of the Central Black Hole}, Astrophys. J. 875~(1) (2019) L6.
\newblock \href {http://arxiv.org/abs/1906.11243} {\path{arXiv:1906.11243}},
  \href {https://doi.org/10.3847/2041-8213/ab1141}
  {\path{doi:10.3847/2041-8213/ab1141}}.

\bibitem{Gralla:2019xty}
S.~E. Gralla, D.~E. Holz, R.~M. Wald, {Black Hole Shadows, Photon Rings, and
  Lensing Rings}, Phys. Rev. D 100~(2) (2019) 024018.
\newblock \href {http://arxiv.org/abs/1906.00873} {\path{arXiv:1906.00873}},
  \href {https://doi.org/10.1103/PhysRevD.100.024018}
  {\path{doi:10.1103/PhysRevD.100.024018}}.

\bibitem{Liebes:1964zz}
S.~Liebes, {Gravitational Lenses}, Phys. Rev. 133 (1964) B835--B844.
\newblock \href {https://doi.org/10.1103/PhysRev.133.B835}
  {\path{doi:10.1103/PhysRev.133.B835}}.

\bibitem{Refsdal:1993kf}
S.~Refsdal, J.~Surdej, {Gravitational lenses}, Rept. Prog. Phys. 57 (1994)
  117--186.
\newblock \href {https://doi.org/10.1088/0034-4885/57/2/001}
  {\path{doi:10.1088/0034-4885/57/2/001}}.

\bibitem{Nzioki:2010nj}
A.~M. Nzioki, P.~K.~S. Dunsby, R.~Goswami, S.~Carloni, {A Geometrical Approach
  to Strong Gravitational Lensing in f(R) Gravity}, Phys. Rev. D83 (2011)
  024030.
\newblock \href {http://arxiv.org/abs/1002.2056} {\path{arXiv:1002.2056}},
  \href {https://doi.org/10.1103/PhysRevD.83.024030}
  {\path{doi:10.1103/PhysRevD.83.024030}}.

\bibitem{kuniyal2018strong}
R.~S. Kuniyal, H.~Nandan, U.~Papnoi, R.~Uniyal, K.~Purohit, Strong lensing and
  observables around 5d myers--perry black hole spacetime, Modern Physics
  Letters A 33~(23) (2018) 1850126.

\bibitem{Uniyal:2018ngj}
R.~Uniyal, H.~Nandan, P.~Jetzer, {Bending angle of light in equatorial plane of
  Kerr–Sen Black Hole}, Phys. Lett. B782 (2018) 185--192.
\newblock \href {http://arxiv.org/abs/1803.04268} {\path{arXiv:1803.04268}},
  \href {https://doi.org/10.1016/j.physletb.2018.05.006}
  {\path{doi:10.1016/j.physletb.2018.05.006}}.

\bibitem{Azreg-Ainou:2017obt}
M.~Azreg-Aïnou, S.~Bahamonde, M.~Jamil, {Strong Gravitational Lensing by a
  Charged Kiselev Black Hole}, Eur. Phys. J. C77~(6) (2017) 414.
\newblock \href {http://arxiv.org/abs/1701.02239} {\path{arXiv:1701.02239}},
  \href {https://doi.org/10.1140/epjc/s10052-017-4969-4}
  {\path{doi:10.1140/epjc/s10052-017-4969-4}}.

\bibitem{Bozza:2001xd}
V.~Bozza, S.~Capozziello, G.~Iovane, G.~Scarpetta, {Strong field limit of black
  hole gravitational lensing}, Gen. Rel. Grav. 33 (2001) 1535--1548.
\newblock \href {http://arxiv.org/abs/gr-qc/0102068}
  {\path{arXiv:gr-qc/0102068}}, \href {https://doi.org/10.1023/A:1012292927358}
  {\path{doi:10.1023/A:1012292927358}}.

\bibitem{Bozza:2002zj}
V.~Bozza, {Gravitational lensing in the strong field limit}, Phys. Rev. D66
  (2002) 103001.
\newblock \href {http://arxiv.org/abs/gr-qc/0208075}
  {\path{arXiv:gr-qc/0208075}}, \href
  {https://doi.org/10.1103/PhysRevD.66.103001}
  {\path{doi:10.1103/PhysRevD.66.103001}}.

\bibitem{Synge:1966okc}
J.~L. Synge, {The Escape of Photons from Gravitationally Intense Stars}, Mon.
  Not. Roy. Astron. Soc. 131~(3) (1966) 463--466.
\newblock \href {https://doi.org/10.1093/mnras/131.3.463}
  {\path{doi:10.1093/mnras/131.3.463}}.

\bibitem{hawking1973black}
S.~Hawking, B.~Carter, J.~M. Bardeen, H.~Gursky, K.~S. Thorne, R.~Ruffini,
  I.~D. Novikov, et~al., Black Holes, Vol.~23, CRC Press, 1973.

\bibitem{Grenzebach:2014fha}
A.~Grenzebach, V.~Perlick, C.~Lämmerzahl, {Photon Regions and Shadows of
  Kerr-Newman-NUT Black Holes with a Cosmological Constant}, Phys. Rev.
  D89~(12) (2014) 124004.
\newblock \href {http://arxiv.org/abs/1403.5234} {\path{arXiv:1403.5234}},
  \href {https://doi.org/10.1103/PhysRevD.89.124004}
  {\path{doi:10.1103/PhysRevD.89.124004}}.

\bibitem{Grenzebach:2015oea}
A.~Grenzebach, V.~Perlick, C.~Lämmerzahl, {Photon Regions and Shadows of
  Accelerated Black Holes}, Int. J. Mod. Phys. D24~(09) (2015) 1542024.
\newblock \href {http://arxiv.org/abs/1503.03036} {\path{arXiv:1503.03036}},
  \href {https://doi.org/10.1142/S0218271815420249}
  {\path{doi:10.1142/S0218271815420249}}.

\bibitem{Johannsen:2015qca}
T.~Johannsen, {Photon Rings around Kerr and Kerr-like Black Holes}, Astrophys.
  J. 777 (2013) 170.
\newblock \href {http://arxiv.org/abs/1501.02814} {\path{arXiv:1501.02814}},
  \href {https://doi.org/10.1088/0004-637X/777/2/170}
  {\path{doi:10.1088/0004-637X/777/2/170}}.

\bibitem{Younsi:2016azx}
Z.~Younsi, A.~Zhidenko, L.~Rezzolla, R.~Konoplya, Y.~Mizuno, {New method for
  shadow calculations: Application to parametrized axisymmetric black holes},
  Phys. Rev. D94~(8) (2016) 084025.
\newblock \href {http://arxiv.org/abs/1607.05767} {\path{arXiv:1607.05767}},
  \href {https://doi.org/10.1103/PhysRevD.94.084025}
  {\path{doi:10.1103/PhysRevD.94.084025}}.

\bibitem{Hioki:2009na}
K.~Hioki, K.-i. Maeda, {Measurement of the Kerr Spin Parameter by Observation
  of a Compact Object's Shadow}, Phys. Rev. D80 (2009) 024042.
\newblock \href {http://arxiv.org/abs/0904.3575} {\path{arXiv:0904.3575}},
  \href {https://doi.org/10.1103/PhysRevD.80.024042}
  {\path{doi:10.1103/PhysRevD.80.024042}}.

\bibitem{Narayan:2019imo}
R.~Narayan, M.~D. Johnson, C.~F. Gammie, {The Shadow of a Spherically Accreting
  Black Hole}, Astrophys. J. 885~(2) (2019) L33.
\newblock \href {http://arxiv.org/abs/1910.02957} {\path{arXiv:1910.02957}},
  \href {https://doi.org/10.3847/2041-8213/ab518c}
  {\path{doi:10.3847/2041-8213/ab518c}}.

\bibitem{Schneider:2018hge}
S.~Schneider, V.~Perlick, {The shadow of a collapsing dark star}, Gen. Rel.
  Grav. 50~(6) (2018) 58.
\newblock \href {http://arxiv.org/abs/1802.04901} {\path{arXiv:1802.04901}},
  \href {https://doi.org/10.1007/s10714-018-2379-z}
  {\path{doi:10.1007/s10714-018-2379-z}}.

\bibitem{Cunha:2016wzk}
P.~V.~P. Cunha, C.~A.~R. Herdeiro, B.~Kleihaus, J.~Kunz, E.~Radu, {Shadows of
  Einstein–dilaton–Gauss–Bonnet black holes}, Phys. Lett. B768 (2017)
  373--379.
\newblock \href {http://arxiv.org/abs/1701.00079} {\path{arXiv:1701.00079}},
  \href {https://doi.org/10.1016/j.physletb.2017.03.020}
  {\path{doi:10.1016/j.physletb.2017.03.020}}.

\bibitem{Vagnozzi:2020quf}
S.~Vagnozzi, C.~Bambi, L.~Visinelli, {Concerns regarding the use of black hole
  shadows as standard rulers}\href {http://arxiv.org/abs/2001.02986}
  {\path{arXiv:2001.02986}}, \href {https://doi.org/10.1088/1361-6382/ab7965}
  {\path{doi:10.1088/1361-6382/ab7965}}.

\bibitem{Tsupko:2019pzg}
O.~{\relax Yu}. Tsupko, Z.~Fan, G.~S. Bisnovatyi-Kogan, {Black hole shadow as a
  standard ruler in cosmology}, Class. Quant. Grav. 37~(6) (2020) 065016.
\newblock \href {http://arxiv.org/abs/1905.10509} {\path{arXiv:1905.10509}},
  \href {https://doi.org/10.1088/1361-6382/ab6f7d}
  {\path{doi:10.1088/1361-6382/ab6f7d}}.

\bibitem{Papnoi:2014aaa}
U.~Papnoi, F.~Atamurotov, S.~G. Ghosh, B.~Ahmedov, {Shadow of five-dimensional
  rotating Myers-Perry black hole}, Phys. Rev. D90~(2) (2014) 024073.
\newblock \href {http://arxiv.org/abs/1407.0834} {\path{arXiv:1407.0834}},
  \href {https://doi.org/10.1103/PhysRevD.90.024073}
  {\path{doi:10.1103/PhysRevD.90.024073}}.

\bibitem{Atamurotov:2015nra}
F.~Atamurotov, B.~Ahmedov, {Optical properties of black hole in the presence of
  plasma: shadow}, Phys. Rev. D92 (2015) 084005.
\newblock \href {http://arxiv.org/abs/1507.08131} {\path{arXiv:1507.08131}},
  \href {https://doi.org/10.1103/PhysRevD.92.084005}
  {\path{doi:10.1103/PhysRevD.92.084005}}.

\bibitem{Perlick:2015vta}
V.~Perlick, O.~{\relax Yu}. Tsupko, G.~S. Bisnovatyi-Kogan, {Influence of a
  plasma on the shadow of a spherically symmetric black hole}, Phys. Rev.
  D92~(10) (2015) 104031.
\newblock \href {http://arxiv.org/abs/1507.04217} {\path{arXiv:1507.04217}},
  \href {https://doi.org/10.1103/PhysRevD.92.104031}
  {\path{doi:10.1103/PhysRevD.92.104031}}.

\bibitem{Konoplya:2019sns}
R.~A. Konoplya, {Shadow of a black hole surrounded by dark matter}, Phys. Lett.
  B795 (2019) 1--6.
\newblock \href {http://arxiv.org/abs/1905.00064} {\path{arXiv:1905.00064}},
  \href {https://doi.org/10.1016/j.physletb.2019.05.043}
  {\path{doi:10.1016/j.physletb.2019.05.043}}.

\bibitem{10.1093/mnras/sty2624}
R.~Shaikh, P.~Kocherlakota, R.~Narayan, P.~S. Joshi, {Shadows of spherically
  symmetric black holes and naked singularities}, MNRAS 482.
\newblock \href {http://arxiv.org/abs/arXiv:1802.08060}
  {\path{arXiv:arXiv:1802.08060}}, \href
  {https://doi.org/10.1093/mnras/sty2624} {\path{doi:10.1093/mnras/sty2624}}.

\bibitem{Jaroszynski:1997bw}
M.~Jaroszynski, A.~Kurpiewski, {Optics near kerr black holes: spectra of
  advection dominated accretion flows}, Astron. Astrophys. 326 (1997) 419.
\newblock \href {http://arxiv.org/abs/astro-ph/9705044}
  {\path{arXiv:astro-ph/9705044}}.

\bibitem{PhysRevD.99.104018}
G.~G.~L. Nashed, S.~Capozziello,
  \href{https://link.aps.org/doi/10.1103/PhysRevD.99.104018}{Charged
  spherically symmetric black holes in $f\mathbf{(}r\mathbf{)}$ gravity and
  their stability analysis}, Phys. Rev. D 99 (2019) 104018.
\newblock \href {https://doi.org/10.1103/PhysRevD.99.104018}
  {\path{doi:10.1103/PhysRevD.99.104018}}.
\newline\urlprefix\url{https://link.aps.org/doi/10.1103/PhysRevD.99.104018}

\bibitem{Garfinkle:1990qj}
D.~Garfinkle, G.~T. Horowitz, A.~Strominger, {Charged black holes in string
  theory}, Phys. Rev. D43 (1991) 3140, [Erratum: Phys. Rev.D45,3888(1992)].
\newblock \href {https://doi.org/10.1103/PhysRevD.43.3140,
  10.1103/PhysRevD.45.3888} {\path{doi:10.1103/PhysRevD.43.3140,
  10.1103/PhysRevD.45.3888}}.

\bibitem{Horowitz:1992jp}
G.~T. Horowitz, {The dark side of string theory: Black holes and black
  strings.}, in: {In *Trieste 1992, Proceedings, String theory and quantum
  gravity '92* 55-99}, 1992.
\newblock \href {http://arxiv.org/abs/hep-th/9210119}
  {\path{arXiv:hep-th/9210119}}.

\bibitem{Dasgupta:2008in}
A.~Dasgupta, H.~Nandan, S.~Kar, {Kinematics of geodesic flows in stringy black
  hole backgrounds}, Phys. Rev. D79 (2009) 124004.
\newblock \href {http://arxiv.org/abs/0809.3074} {\path{arXiv:0809.3074}},
  \href {https://doi.org/10.1103/PhysRevD.79.124004}
  {\path{doi:10.1103/PhysRevD.79.124004}}.

\bibitem{Gerard:1999wd}
J.~Gerard, S.~Pireaux, {The Observable light deflection angle}\href
  {http://arxiv.org/abs/gr-qc/9907034} {\path{arXiv:gr-qc/9907034}}.

\bibitem{PhysRevD.87.107501}
C.~Bambi, Can the supermassive objects at the centers of galaxies be
  traversable wormholes?, Phys. Rev. D 87.
\newblock \href {https://doi.org/10.1103/PhysRevD.87.107501}
  {\path{doi:10.1103/PhysRevD.87.107501}}.

\bibitem{shreyas_bapat_2019_3526198}
S.~Bapat, et~al., einsteinpy/einsteinpy: Einsteinpy 0.2.1 (2019).
\newblock \href {https://doi.org/10.5281/zenodo.3526198}
  {\path{doi:10.5281/zenodo.3526198}}.

\end{thebibliography}
\end{document}